# Beyond Technical Debt: How AI-Assisted Development Creates Comprehension Debt in Resource-Constrained Indie Teams

**Yujie Zhang**
MC232 Masters of Animation, Games & Interactivity
RMIT University
s3918396@rmit.student.edu.au

## Abstract

Junior indie game developers in distributed teams face persistent challenges in technical debt, coordination, and burnout. Traditional methodologies (GDLC, Scrum) assume contexts unavailable to part-time, distributed novice teams. Frameworks addressing AI-augmented production for resource-constrained teams are lacking.

This practice-based exploratory study presents the CIGDI (Co-Intelligence Game Development Ideation) Framework, which emerged from developing *The Worm's Memoirs*, a 2D point-and-click narrative game. Through reflective practice methodology (Schön, 1983) and autoethnographic documentation, we reconstructed patterns from three months of development by a three-person distributed team. The study examines how AI tools can support production challenges when traditional methodologies are inaccessible, analyzing multiple data sources: Jira task management (N=157 tasks), Miro visual documentation boards (N=13+), GitHub commits (N=333), team reflection sessions (N=8), and emotional documentation.

The framework consists of seven iterative stages (Research, Ideation, Prototyping, Playtest, Review, Action, and Combination) structured around two key human-in-the-loop decision points: Priority Criteria and Timeboxing. The framework emphasizes human-led decision-making, with AI supporting production tasks (documentation, code scaffolding, research access) while humans maintain creative control and verification. Retrospective analysis revealed both benefits—knowledge access democratization, reduced cognitive load—and significant challenges. Notably, we identified **"comprehension debt"** as a novel form of technical debt: AI helped us build more sophisticated systems than our skill level could independently create, but our insufficient expertise made those systems difficult to maintain or modify. This paradox—possessing working systems we incompletely understood—created fragility and AI dependency distinct from traditional code quality debt.

This work contributes an alternative production approach for resource-constrained indie teams, offering practical guidance for integrating AI tools with appropriate ethical boundaries and verification protocols. The research demonstrates the value of honest, messy practice-based documentation and identifies critical questions about whether AI assistance constitutes a learning ladder or dependency trap for skill development.

# Introduction

The indie game development industry faces persistent challenges in production management, team coordination, and developer wellbeing. Research shows that 76% of game developers experience crunch culture (Peticca-Harris et al., 2015; Weststar et al., 2020). Analysis of 927 problems from 200 game development postmortems revealed that most root causes relate to people rather than technology (Politowski et al., 2021). Small teams also suffer from estimation errors (Jørgensen et al., 2004) due to the planning fallacy (Kahneman & Tversky, 1979). These patterns are particularly acute for junior developers working in distributed, part-time teams with limited resources.

Our three-person team (artist, sound designer, programmer) developed *The Worm's Memoirs* over three months part-time with zero budget. We faced technical challenges (point-and-click architecture expertise gaps, technical debt accumulation; Politowski et al., 2021), coordination issues (asynchronous collaboration limits, communication barriers; Jackson et al., 2024; Caravella et al., 2022; Mok et al., 2023), and human factors (stress, burnout; Washburn et al., 2016).

Traditional methodologies such as GDLC and Scrum provide valuable structures (Siahaan et al., 2023). However, these approaches assume full-time, co-located teams. For junior developers in distributed, part-time teams, these assumptions don't hold. Emerging AI tools offer potential to reduce cognitive load (Noy & Zhang, 2023), yet there is limited research on structuring AI integration for novice developers with explicit ethical boundaries. Particularly lacking are frameworks that balance AI assistance for production tasks while avoiding over-reliance risks (Perry et al., 2023; Kazemitabaar et al., 2023).

This study addresses the following research question:

**RQ: How can AI tools support novice indie game developers in managing production challenges when traditional methodologies are inaccessible?**

Specifically: (1) What production challenges do distributed junior teams face? (2) How can AI integration support production while maintaining human creative control? (3) What risks and benefits emerge from AI-augmented development in resource-constrained contexts?

# Related Work

**Technical Production Challenges**

Technical debt and insufficient expertise are significant concerns for novice indie developers. Politowski et al. (2021) analyzed 927 problems from 200 postmortems, finding most root causes related to people rather than technologies—specifically, knowledge gaps between developers. Small teams (≤20) face obstacles (37%), schedule issues (25%), and process challenges (24%), with experts providing 20-30% less realistic estimates due to planning fallacy (Jørgensen et al., 2004; Kahneman & Tversky, 1979). Washburn et al. (2016) documented similar patterns. Our three-person team lacked point-and-click architecture expertise, accumulating technical debt documented across 333 GitHub commits showing constant refactoring cycles.

### Team Coordination Issues

Distributed collaboration presents well-documented challenges. Caravella et al. (2022) documented communication barriers in remote game development with productivity and mental health impacts. Mok et al. (2023) analyzed 20 million meetings at Microsoft, revealing asymmetric burdens where distributed members experience scheduling conflicts. Jackson et al. (2024) examined co-creation patterns in remote software teams, finding asynchronous collaboration limits real-time problem-solving. For small indie teams lacking documentation practices, these challenges intensify—knowledge gaps create miscommunication, and documentation becomes simultaneously essential and frustrating (Colby & Colby, 2019).

### Human Factors & Wellbeing

Stress, burnout, and crunch culture represent persistent concerns in game development, with 76% of developers experiencing crunch (Peticca-Harris et al., 2015; Weststar et al., 2020). Mendes and Queirós (2022) documented moderate burnout levels across 193 game developers. Graziotin et al. (2017) demonstrated that developer emotions directly impact code quality and decision-making—frustration leads to impulsive decisions and reduced problem-solving ability. Cote and Harris (2023) analyzed how "good crunch" discourse perpetuates unsustainable practices even in voluntarist contexts.

### Human-AI Collaboration

Recent research documents significant challenges when novice programmers use AI code generation tools. Students develop an "illusion of competence," believing they understand AI-generated code when they do not (Prather et al., 2024). Over-reliance patterns emerge where students abandon traditional learning resources for sole AI dependence without improved outcomes (Xue et al., 2024). Studies reveal problematic interaction patterns including "shepherding" and "drifting," with these behaviors negatively correlating with performance (Prather et al., 2023). This research establishes that AI tools can undermine skill development despite improving task completion rates.

### Positioning: What's Missing

Traditional methodologies assume full-time commitment, co-located teams, and experienced leadership—conditions inaccessible to part-time, distributed, novice developers. Emerging AI tools offer potential to reduce cognitive load and democratize knowledge access (Noy &

Zhang, 2023), yet frameworks specifically addressing AI integration for resource-constrained indie teams with appropriate ethical boundaries and verification protocols remain limited.

# Methodology

### Research Approach

This study employs practice-based research drawing on Schön's (1983) reflective practice where knowledge emerges from systematic inquiry into one's own practice. We employed both reflection-in-action (documenting during development) and reflection-on-action (retrospective pattern analysis). We also use autoethnographic documentation (Ellis et al., 2011) to capture emotional and experiential dimensions.

### Project Context

The CIGDI framework emerged from developing *The Worm's Memoirs*, a 2D point-and-click narrative game about childhood trauma. The three-person team (artist, narrative designer/sound designer, programmer) worked over three months (February-April 2025), part-time (10-15 hours/week), distributed across time zones, with zero budget. This project was selected because its constraints represent common conditions in indie game development.

### Data Collection

We collected multiple data sources retrospectively and during development:

- **Jira Tasks**: N=157 tasks documenting work items, priorities, time estimates
- **GitHub Commits**: N=333 commits showing code evolution, refactoring patterns
- **Miro Boards**: N=13+ visual documentation boards for sprint planning, retrospectives
- **Team Reflection Sessions**: N=8 recorded sessions analyzing process, challenges
- **Microsoft Teams Communication**: Messages, calls documenting coordination
- **Emotional Documentation**: Memes, diary entries, frustration logs

### Analysis Method

We employed **retrospective pattern reconstruction** through thematic analysis of development artifacts. Rather than pre-defining framework components, we allowed patterns to emerge from documented practice. The researcher (as participant-developer) analyzed commits, task descriptions, and reflection sessions to identify recurring production challenges, decision points, and AI tool usage patterns. These patterns were iteratively refined through team discussions and validated against artifact evidence.

### Positionality & Limitations

The researcher served as both framework creator and primary user (programmer role), creating potential bias. As a junior developer, my technical limitations shaped both challenges faced and solutions attempted. The three-month timeline, zero budget, and part-time constraints were genuine—not experimental conditions—reflecting authentic indie development realities.

This is exploratory documentation of ONE team's experience, not validated methodology. Findings represent transferable insights for similar contexts (small, distributed, junior teams) rather than universal claims. We explicitly acknowledge messiness, failures, and ongoing uncertainty about long-term impacts.

# Framework Overview

The CIGDI framework consists of seven iterative stages organized around two critical human-in-the-loop decision points:

**Core Stages**: 1. **Research (AI-Assisted)**: Market analysis, technical feasibility, reference gathering 2. **Ideation**: Concept generation, brainstorming, constraint identification 3. **Prototyping (AI-Assisted)**: Rapid implementation, code scaffolding, asset creation 4. **Playtest**: User feedback collection, observation, documentation 5. **Review & Analysis**: Pattern identification, problem diagnosis, data synthesis 6. **Action Planning**: Task breakdown, resource allocation, timeline estimation 7. **Combination/Iteration**: Integration of changes, next cycle preparation

**Decision Points**: - **Priority Criteria**: Human-defined values for feature selection, scope decisions - **Timeboxing**: Human-set deadlines triggering review cycles, preventing endless iteration

**AI Tool Integration Principles**: - AI for **production support** (documentation, research access, code scaffolding) - Humans for **creative decisions** (narrative, art direction, core mechanics) - Mandatory **verification protocols** for AI-generated code - Explicit **ethical boundaries** preventing AI use in core creative domains

# Findings

### Benefits of AI-Augmented Development

**Knowledge Access Democratization**: AI tools provided instant access to technical knowledge previously requiring years of experience. When facing point-and-click architecture challenges, AI explanations of state machines and event systems accelerated learning compared to traditional documentation.

**Reduced Cognitive Load**: For routine tasks (boilerplate code, documentation formatting, API queries), AI assistance freed mental resources for creative problem-solving. The programmer could focus on game logic rather than syntax lookup.

**Accelerated Prototyping**: AI-generated code scaffolds enabled rapid iteration. What traditionally required days of implementation could be prototyped in hours, enabling more experimental design approaches.

### Challenges & Risks: Comprehension Debt

**The Core Problem**: AI assistance enabled us to build systems more sophisticated than our skill level could independently create. We possessed working code we incompletely understood. When bugs emerged or modifications were needed, we lacked the conceptual foundation to debug or extend these systems confidently.

This differs from traditional technical debt (code quality issues, shortcuts). **Comprehension debt** represents a knowledge gap: the system works, but the team cannot maintain it without AI assistance. This creates:

- **Fragility**: Systems break when modified because developers don't understand internal logic
- **AI Dependency**: Debugging requires returning to AI tools, reinforcing reliance
- **Skill Stagnation**: Copying AI solutions prevents developing problem-solving expertise
- **False Confidence**: Working prototypes mask underlying incompetence

**Documentation Challenges**: AI-generated code often lacked inline comments or clear naming. When the programmer returned to code weeks later, comprehension required AI assistance to explain our own codebase.

**Verification Burden**: The "Trust But Verify" protocol created overhead. Every AI suggestion required testing, but insufficient expertise made verification incomplete. We caught obvious errors but missed subtle architectural problems.

### Emotional & Wellbeing Impacts

Development oscillated between productivity euphoria and incompetence anxiety. AI tools provided dopamine hits from rapid progress, but subsequent debugging failures exposed skill gaps, creating emotional volatility. The team documented frustration through memes and diary entries, revealing a complex relationship with AI assistance—simultaneously empowering and undermining.

## Discussion

### Comprehension Debt: A New Category of Technical Debt

Kruchten et al. (2012) defined technical debt as shortcuts compromising future maintainability. Comprehension debt extends this: AI-assisted development creates systems exceeding the team's expertise, generating future maintenance challenges not from poor code quality but from insufficient understanding. This suggests AI tools may enable a "construction-comprehension gap" where building outpaces learning.

### Learning Ladder or Dependency Trap?

The critical question remains unresolved: Does AI assistance scaffold skill development (Vygotsky's Zone of Proximal Development) or create permanent dependency? Our three-month timeline is too short to determine long-term impacts. We observed both patterns:

- **Scaffolding Evidence**: Some AI explanations improved

- conceptual understanding
  - **Dependency Evidence**: Repeated reliance on AI for similar problems without retained learning

This tension aligns with educational research (Prather et al., 2024; Xue et al., 2024) showing AI tools can undermine learning despite improving task performance.

### Implications for Resource-Constrained Teams

For indie developers with limited time and budgets, comprehension debt may represent an acceptable trade-off. Shipping a working prototype with maintenance risks may be preferable to never completing the project. However, this decision should be conscious, not accidental.

The CIGDI framework attempts to manage this trade-off through explicit verification protocols and ethical boundaries. Whether these safeguards suffice remains uncertain.

### Limitations & Future Research

This single-case, three-month study with practitioner-researcher bias cannot determine long-term impacts. Future research should:

- Longitudinal studies tracking skill development with/without AI assistance
- Comparative studies across different team sizes, experience levels
- Measurement frameworks for comprehension debt quantification
- Intervention studies testing different verification protocols

# Conclusion

This practice-based study documents how AI-assisted development creates comprehension debt—a novel form of technical debt where teams build systems exceeding their expertise. The CIGDI framework represents an exploratory approach for resource-constrained indie teams, emphasizing human-led decision-making with AI production support.

Key contributions include: 1. Identification and documentation of comprehension debt as distinct from traditional technical debt 2. A framework balancing AI assistance with human creative control and verification 3. Honest documentation of both benefits and failures in AI-augmented development 4. Demonstration that practice-based research valuing messiness over polish provides valuable insights

For junior indie developers, this work offers practical guidance for integrating AI tools with appropriate skepticism, verification protocols, and ethical boundaries that preserve creative agency. For researchers, it suggests that practice-based documentation valuing honesty over polish provides valuable insights that traditional, "clean" studies may not capture. For the game industry, it demonstrates that alternatives exist when traditional methodologies do not fit. It also highlights the critical, unresolved questions regarding AI's long-term impact on developer skill formation, creative autonomy, and sustainable work practices.